\documentclass[%
 reprint,
 amsmath, amssymb,
 aps,
prl,
]{revtex4-1}

\usepackage{graphicx}
\usepackage{dcolumn}
\usepackage{bm}
\usepackage{hyperref}
\usepackage{subfigure}
\usepackage{siunitx}
\usepackage{verbatim}
\usepackage{amsmath}
\usepackage{amssymb}
\usepackage{bbold}
\usepackage{chemformula}

\usepackage{url}

\begin{document}

\preprint{APS/123-QED}

\title{
	Anisotropic van der Waals Dispersion Forces in Polymers:\\ 
	Structural Symmetry Breaking Leads to Enhanced Conformational Search
	}

\author{Mario Galante}
\author{Alexandre Tkatchenko}%
\affiliation{%
Department of Physics and Materials Science, University of Luxembourg, L-1511 Luxembourg City, Luxembourg
}%

\date{\today}
\begin{abstract}
	The modeling of conformations and dynamics of (bio)polymers is of primary importance for understanding physicochemical properties of soft matter. 
	Although short-range interactions such as covalent and hydrogen bonding control the local arrangement of polymers, non-covalent interactions play a dominant role in determining the global conformations.
	Here we focus on how the inclusion of many-body effects in van der Waals (vdW) dispersion influence the molecular dynamics of model polymers and a poly-alanine chain studied with semiempirical quantum mechanics.
	We show that delocalized force contributions are key to explore the conformational landscape, as they induce an anisotropic polarization response which efficiently guides the conformation towards globally optimized structures. 
	This is in contrast with the commonly used pair-wise interatomic Lennard-Jones-like potentials, which result in structures that optimize local regularity rather than global symmetries. The many-body vdW formalism results in a significant reduction in the roughness of the energy landscape and drives the conformational search towards geometries that are consistent with global spatial symmetries.

\end{abstract}

                              %
\maketitle

The prediction of structure formation in polymers under realistic thermodynamical conditions is a challenging problem that is relevant for fields ranging from medicine and biology to physics and engineering.
For example, the folding of proteins into their secondary structure underpins the regulation of many functions of cellular activity \cite{Bu2011}.
In addition, the ubiquitous presence of polymeric materials in modern applications is due to their variety of macroscopic properties which are in turn dictated by polymer morphology and its dynamics \cite{Adhikari2004, Dersch2005, Ulery2011}.
However, conformational stabilities are easily undermined by thermal fluctuations and the solvent environment, therefore the knowledge of significant portions of the potential energy surface (PES) of the system become essential \cite{Napolitano2015, Fraser2009, Uversky2009}.
Models based on phenomenological definitions of inter-fragment interactions \cite{Souza2021} have consistently proved to be invaluable tools for accurate predictions of dynamical properties in biological systems, as they allow to probe time and length scales inaccessible to atomistic simulations.
Nevertheless, modern accurate electronic structure calculations have become feasible for systems of biological relevance.
The analysis of electron correlations at such scales can therefore shed new light on mechanisms driving dynamical processes, and therefore be inspiration for the further development of accurate force fields~\cite{Unke2021a, Unke2021}.



The determination of an appropriate first-principles description for the molecular interactions that are relevant for large and relatively flexible systems has thus become a timely and challenging issue.
The general physical model \cite{Tran2020, Abeln2019} includes bonded interactions together with non-covalent potentials to account for electrostatics and van der Waals (vdW) dispersion interactions.
From a physical perspective, these are the result of the interaction between instantaneous charge fluctuations \cite{Hermann2017, Stoehr2019} and play a dominant role in determining the global arrangement because of their strikingly slow spatial decay.
Traditionally, they are included by means of the pair-wise (PW) approximation, where the energy decays strictly with the $6$th power of the interatomic distance and its strength is dictated by the local chemical environments \cite{Tkatchenko2009, Sato2009, Grimme2010}.
However, recent studies \cite{Dobson2014, Ambrosetti2016, Stoehr2019, Hauseux2020} have shown that the poor versatility that derives from the lack of many-body effects makes the PW approximation not suitable to adequately capture the complexity of vdW interactions in a variety of systems. However, the implications of noncovalent many-body interactions on molecular dynamics have not been studied.


In this Letter we focus on the qualitative features of the conformational search for PW and many-body dispersion (MBD) potentials by means of molecular dynamics (MD) simulations for a model carbon-based polymer.
We show that the many-body character of vdW forces enables one to capture anisotropic polarizability effects, which consistently drive the molecular structure to optimize the global symmetry of the conformation according to geometrical constraints.
On the contrary, the shorter range of the forces within the PW approximations, together with their independence on the global atomic configuration, yields conformations that are less compact and an increased roughness of the PES. 
We then present the results of MD simulations for an \ch{Ala_{15}} polypeptide where the electronic interactions are treated with semi-empirical quantum mechanics, which confirms the trends observed for the model polymer system.

We begin by considering a toy model carbon chain represented by the model Hamiltonian
\begin{eqnarray}
	\mathcal{H} &&= \sum_{i=1}^{N-1} k\Big( \big|\big| \boldsymbol{R}_{i}-\boldsymbol{R}_{i+1} \big|\big| - R_0\Big)^2 \\
	&&+ \sum_{i=1}^{N} \sum_{j\neq i,i\pm 1} a \ \text{exp}\Big\{-b \big|\big|\boldsymbol{R}_{i}-\boldsymbol{R}_{j}\big|\big|\Big\} + E_{\rm{vdW}}, \nonumber
\end{eqnarray}
with $\boldsymbol{R}_i$ representing the position of the $i$th carbon atom, $R_0=$\SI{1.52}{\angstrom}, $k=39.57$ a.u.~, $a=1878.38$ a.u.~and $b=5.49$ a.u.~.
The first term represents an harmonic covalent bond between the nearest-neighbours, while the second summation introduces a Born-Mayer repulsive potential between non-nearest-neighbours that prevents artificial agglomeration of the polymer. 
We note that the parameters were chosen in order to ensure a solid backbone, however they do not aim to faithfully represent the chemical environment.
In a similar spirit, we choose here to entirely neglect bond angle and dihedral constraints in order to further reduce the number of degrees of freedom and favours a qualitative comparison of the PES resulting from different formulations for the vdW energy, $E_{\rm{vdW}}$.
These assumptions are therefore instrumental to detect the signatures characteristic of MBD in a system of reduced complexity, which then enables targeted analysis in realistic simulation conditions.
Dispersion interactions can be generally quantified by means of the adiabatic connection fluctuation-dissipation (ACFD) formula~\cite{Hermann2017, Tkatchenko2013},
\begin{eqnarray}\label{ACFD}
	E_\text{vdW}^\textrm{MBD} &&= \frac{\hbar}{2\pi} \int_0^\infty du \int_0^1 d\lambda \int \int d\boldsymbol{r}d\boldsymbol{r}^\prime\\
	&&\times \text{tr}\Big[ \big( \boldsymbol{\alpha}_\lambda (\boldsymbol{r}, \boldsymbol{r}^\prime, iu) - \boldsymbol{\alpha}_0 (\boldsymbol{r}, \boldsymbol{r}^\prime, iu) \big) \boldsymbol{T}(\boldsymbol{r}, \boldsymbol{r}^\prime) \Big], \nonumber
\end{eqnarray}
where $\boldsymbol{r}$, $\boldsymbol{r}^\prime$ are spatial variables, $u$ is the frequency of the electric field, tr denotes the trace of the polarizability tensors, $\boldsymbol{\alpha}$, and the integral over the coupling strength, $\lambda$, expresses the adiabatic switching-on of the electron-electron interaction.
Here we approximate the ACFD equation within the many-body dispersion approach \cite{Tkatchenko2012, Tkatchenko2013} and we define the coupling tensor, $\boldsymbol{T}$, as the dipole-dipole potential corresponding to two Gaussian charge distributions.
The polarizability tensor of the system corresponding to uncorrelated atoms, $\boldsymbol{\alpha}_0$, can be written as the sum of the polarizabilities localized at each atomic site,
\begin{equation}\label{alpha:PW}
	\boldsymbol{\alpha}_\text{0} (\boldsymbol{r}, \boldsymbol{r}^\prime) \approx \sum_i  \delta^3(\boldsymbol{r}-\boldsymbol{R}_i) \delta^3(\boldsymbol{r}-\boldsymbol{r}^\prime) \mathbb{1}_3\alpha_i^\text{TS},
\end{equation}
where $\alpha_i^{\rm{TS}}$ accounts for all short range correlations and is here taken as the isotropic atomic polarizability for a generic carbon calculated within the Tkatchenko-Scheffler (TS) scheme \cite{Tkatchenko2009} , $\alpha^\text{TS}_\text{C}=10.44$ a.u.
The polarizability of the interacting system, $\boldsymbol{\alpha}_\lambda$, can then be calculated through the Dyson-like screening equation,
\begin{equation}\label{Dyson}
	\boldsymbol{\alpha}_\lambda = 
	\sum_{n=0}^\infty \langle \boldsymbol{\alpha}_\text{0} (-\lambda \boldsymbol{T} \boldsymbol{\alpha}_\text{0})^n \rangle ,
\end{equation}
which allows to construct a non-diagonal tensor starting from the local atomic polarizabilites.
We remark that the solution of Eqs.~(\ref{ACFD}) and (\ref{Dyson}) within these approximations corresponds to solving the Schr\"odinger equation for a system of quantum harmonic oscillators centered at the atomic sites and coupled via the dipole-dipole potential \cite{Hermann2017}.
Therefore, the MBD formalism accounts for the energy of global charge oscillations expressed through linear combinations of the atomic dipolar displacement vectors.
In other terms, the MBD energy is the sum of the interaction energies of the $3N$ normal modes, with $N$ being the total number of oscillators.
In contrast, the PW approximation corresponds to truncating the series of Eq.~(\ref{Dyson}) to the second order and considers only unscreened dipole-dipole interactions,
\begin{equation}\label{PWenergy}
	E^\text{PW}_\text{vdW} = -\frac{1}{2} \sum_{i\neq j} f_\text{damp}(R_{ij})\frac{C_{6,ij}}{R_{ij}^6},
\end{equation}
where $f_\text{damp}$ is a Fermi-type damping function that eliminates short range contributions and $C_{6,ij} = \frac{3}{\pi} \hbar \int du \alpha^\text{TS}_i (iu) \alpha^\text{TS}_j (iu)$ .

\begin{figure}[!t]
	\centering
	\includegraphics[width=0.49\textwidth]{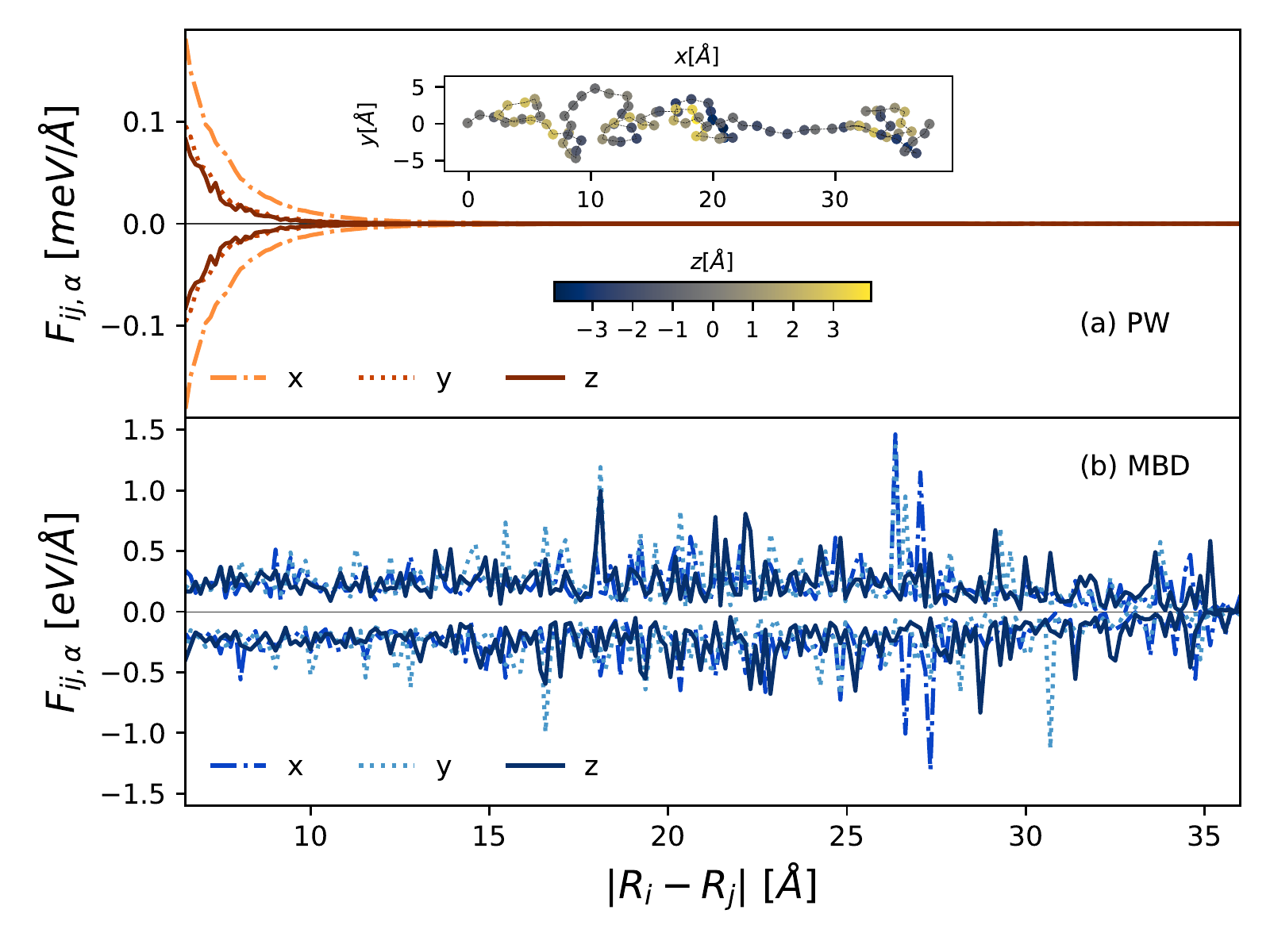}
	\caption{ (Color Online) Dependence of the averaged pair-wise (a) and many-body (b) force components for the interatomic distance for a conformation obtained through molecular dynamics for a $100$ atoms chain (shown in the inset).
	}
	\label{Fig1}
\end{figure}

We begin by analyzing the differences between the long range behavior of the components of the PW and MBD forces for a representative conformation of a 100 atoms chain extracted from a MD trajectory [Fig.~\ref{Fig1}].
For simplicity, we constrain the atoms at each edge of the chain to oscillate around their equilibrium position through a quadratic potential.
The PW force between a given pair of atoms can be calculated analytically from Eq.~\ref{PWenergy} and therefore follows by construction the $R^{-7}$ power law, becoming negligible within \SI{15}{\angstrom}.
On the contrary, the MBD force is the result of the competition between all normal oscillation modes of the coupled dipoles.
Here we numerically calculate the contribution acting on the $i$-th site coming from the $j$-th atom by using the formula,
\begin{equation}
	F_{ij,\alpha}^\textrm{MBD} = -\frac{\partial}{\partial R_i^\alpha} \sum_{m=1}^{3N} \left[ \varepsilon_m \chi_j^m \right], \quad
	\chi_j^m=\sum_{\alpha=x,y,z} \big(\chi_{j,\alpha}^m\big)^2
\end{equation}
where $\varepsilon_m$ is the eigenvalue of the $m$-th MBD mode, and $\chi_{j,\alpha}^m$ is the $\alpha=x,y,z$ component of the displacement vector of the $j$-th dipole within the same mode.
From panel (b) it is evident that the average contribution to MBD forces is entropically scattered throughout the whole distance range and no clear decay rate can be singled out.
This can be attributed to the elevated degree of delocalization of the MBD eigenmodes, as shown in Ref.~\cite{Ambrosetti2016}.
Moreover, the persistence of MBD interactions for long distances has already been reported in model and realistic systems \cite{Haseux2022, Stoehr2019}, and their structure dependence will be analyzed in future publications.
Furthermore, there is no evident relation between the MBD force components and the (cylindrical-like) spatial geometry of the conformation.
This is in contrast with the PW case, where the longitudinal ($x$) component is more pronounced than the others.
We also remark that the typical contribution at a given distance is three orders of magnitude higher for MBD than for PW forces, although in both cases contributions of opposite sign balance to yield an average atomic force of $\approx \SI{0.01}{\electronvolt/\angstrom}$.
This can be associated to the interplay between the contributions due to the $3N$ normal modes, consistently with the analysis of the atomic force response presented in Ref.~\cite{Haseux2021}.

\begin{figure}[!t]
	\centering
	\includegraphics[width=0.49\textwidth]{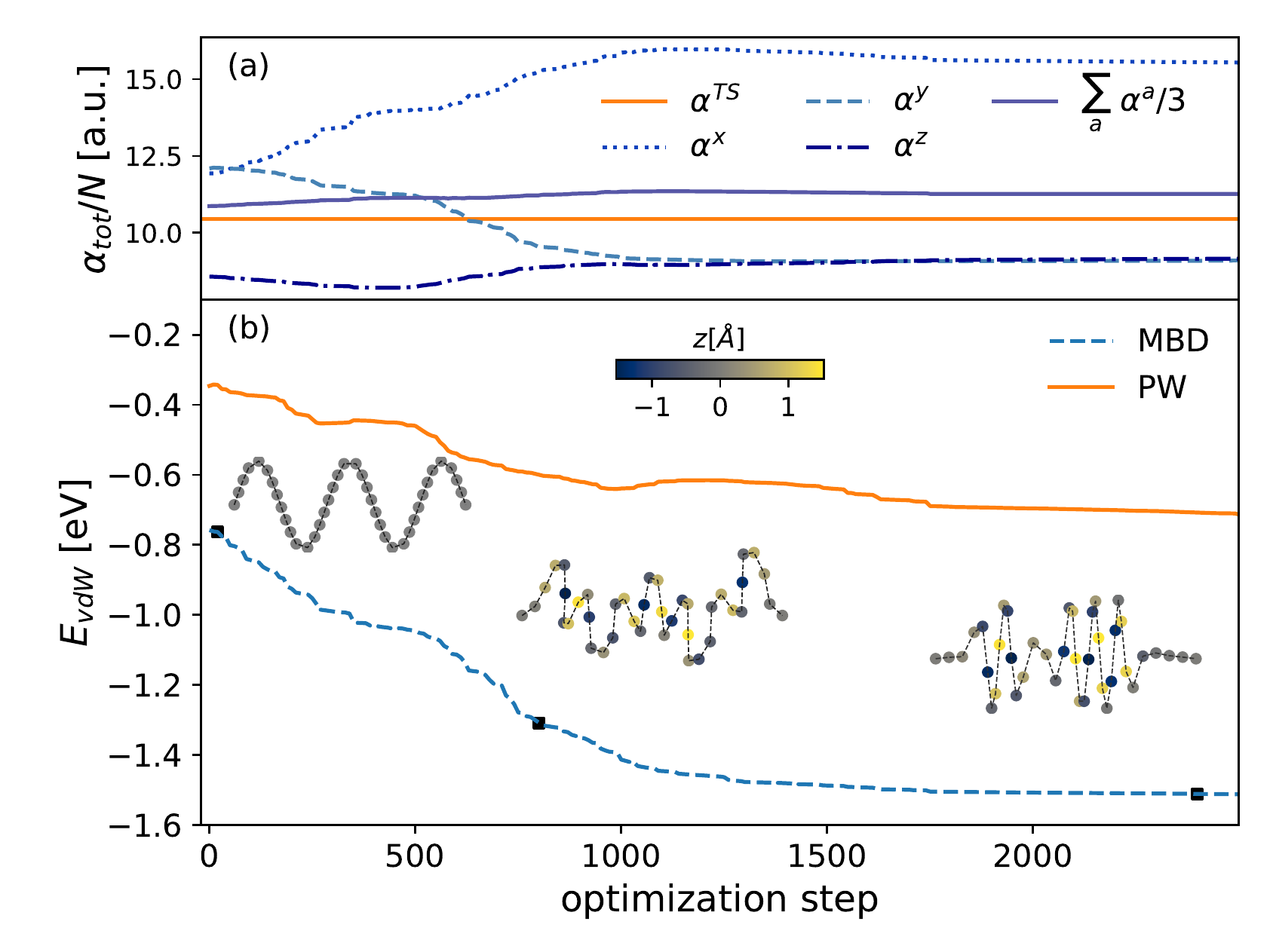}
	\caption{(Color Online) Evolution of the average atomic polarizability, $\alpha$, (top) and of the vdW energy (bottom) along an MBD optimization path.
	Panel (a): the orange solid line represents the input atomic polarizability, $\alpha_C^\text{TS}$, while the dotted, dashed and dot-dashed lines correspond to the $x$, $y$ and $z$ components of the average atomic polarizability, respectively.
	Panel (b): the dashed blue (solid orange) curve represents the trend of the MBD (PW) energy with the progression of the optimization.
	The black squares denote the vdW energy	and optimization step of the remaining three conformations displayed.
	}\label{Fig2}
\end{figure}
We further analyze this by evaluating the polarizability tensor by means of the Dyson-like equation of Eq.~(\ref{Dyson}) for $\lambda=1$ (its trace corresponding to experimentally measured polarizability).
In Fig.~\ref{Fig2} (a) we display the average atomic polarizability along the Cartesian directions,
\begin{equation}
	\alpha_\text{tot}^a = N^{-1}\sum_{ij}\sum_b \alpha_{\lambda=1}^{ab} (\boldsymbol{R}_i, \boldsymbol{R}_j),
\end{equation}
calculated for a series of configurations for a $40$ atoms chain obtained through the Newton-Raphson optimization with respect to the MBD energy of an input two dimensional conformation, while maintaining fixed the coordinates of the edge atoms.
It is clear that the screened atomic total polarizabilty remains mostly constant, similarly to the PW atomic polarizability that is fixed by definition to $\alpha^\textrm{TS}_\textrm{C}$. 
However, the components of the total screened polarizability gradually change to reproduce the the cylindrical-like symmetry imposed by the constrain on the edge atoms, with the longitudinal ($x$) component growing dominant with respect to the others.
We remark that this is analogous to the asymmetry of the force components observed in Fig.~\ref{Fig1}, and it is evident that the anisotropic character of MBD polarizabilities is responsible for driving structures towards three-dimensional conformations.
In fact, the change in the symmetry of the total polarizability goes on equal footing with the lowering of the MBD energy, as shown in Fig.~\ref{Fig2} (b).
For comparison, we display the PW energy calculated for the same structures.
We find that the energy gain along the sequence is significantly less pronounced for PW than for MBD, with the trend of the former displaying two local minima.
This behaviour hints to the fact that MBD forces may favour a more efficient conformational optimization than their PW counterpart.
We note that the increase of the efficiency of conformational seaches, measured in terms of the decrease of energy landscape roughness, has already been linked with the increase of interaction range in the context of pair-wise potentials \cite{Hoare1976, Hoare1983, Doye1996, Rechtsman2005, Rechtsman2006}.


\begin{figure}[!t]
	\centering
	\includegraphics[width=0.49\textwidth]{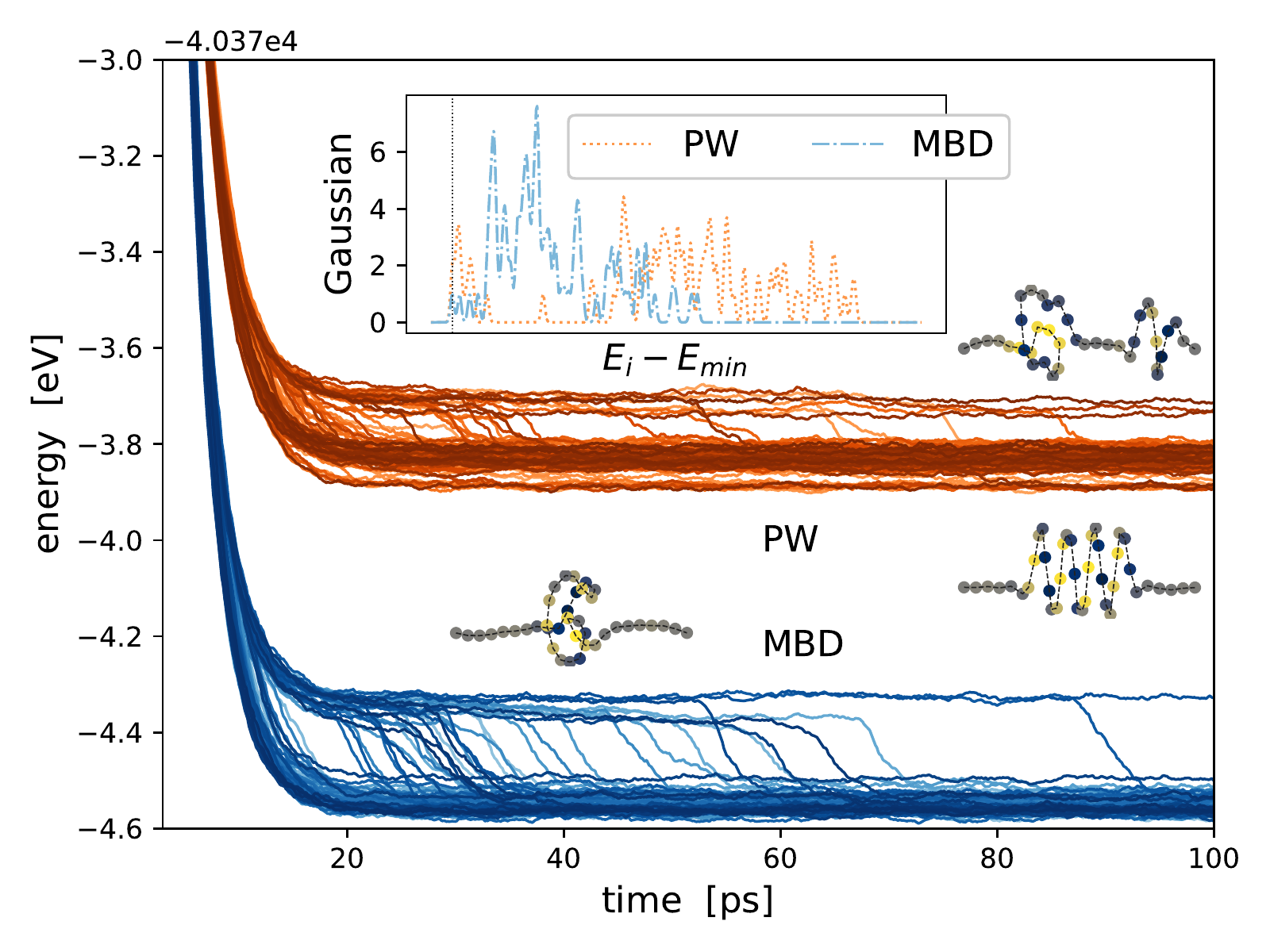}
	\caption{(Color Online) Results of MD simulations (100 per vdW model) at \SI{5}{\kelvin} for a $40$ atoms chain starting from the 2D conformation defined by $n=5$ and $f=2$.
	Orange (blue) lines represent the evolution of the total energy with time for the PW (MBD) approximation of the vdW energy.
	The structures on the right represent the steady-state PW conformations with the highest (top) and the lowest (bottom) energy, while the structure on the left is a typical MBD steady-state structure.
	The inset shows the Gaussian distributions centered on steady-state energies for a each vdW model, rescaled by $E_\text{min}=\text{min}\{E_i\}$ and normalized so that they peak at $1$.
	}
	\label{Fig3}
\end{figure}

To confirm our hypothesis, we performed $100$ MD simulations for each vdW model, all run at \SI{5}{\kelvin} and starting from the same 2D conformation of the $40$ atom polymer displayed in Fig.~\ref{Fig2}.
In Fig.~\ref{Fig3} we display the time evolution up to \SI{100}{\pico\second} of the total energy for all MD trajectories, with the inset showing the Gaussian-broadened distribution of the steady-state energies.
We exclude from the latter the 5 highest energies as they are not statistically representative.
It is clear that PW trajectories yield a large variety of steady-state energies, with the lowest distinctly separated from the others, while MBD energies are more condensed towards the lowest minimum, $E_\text{min}$.
To quantify this difference we calculated the roughness of the potential energy landscape, defined as the root-mean-squared energy distance from $E_\text{min}$, $\langle (E-\text{min}\{E\})^2\rangle^{1/2}$ \cite{Milanesi2012}, which gives $\sim 137 \ k_\text{B} T$ and $\sim 63 \ k_\text{B}T$ for PW and MBD, respectively.
Moreover, 19 of the MBD trajectories undergo a transition after the initial transient ($t\gtrsim$\SI{20}{\pico\second}), while this occurs only in 10 cases for PW.
Although the height of the energy barriers is roughly similar ($\approx$\SI{10}{\milli\electronvolt}), MBD transitions have average longer duration ($\approx$\SI{34}{\pico\second} against \SI{23}{\pico\second}) and result in a larger energy decrease ($\approx$\SI{0.21}{\electronvolt} against \SI{0.15}{\electronvolt}).
Therefore, we can conclude that the introduction of anisotropic polarizabilities in the model polymer results in a two-fold decrease of PES roughness and enchances the likelyhood of structural transitions towards lower energy local minima.

\begin{figure}[!t]
	\centering
	\includegraphics[width=0.49\textwidth]{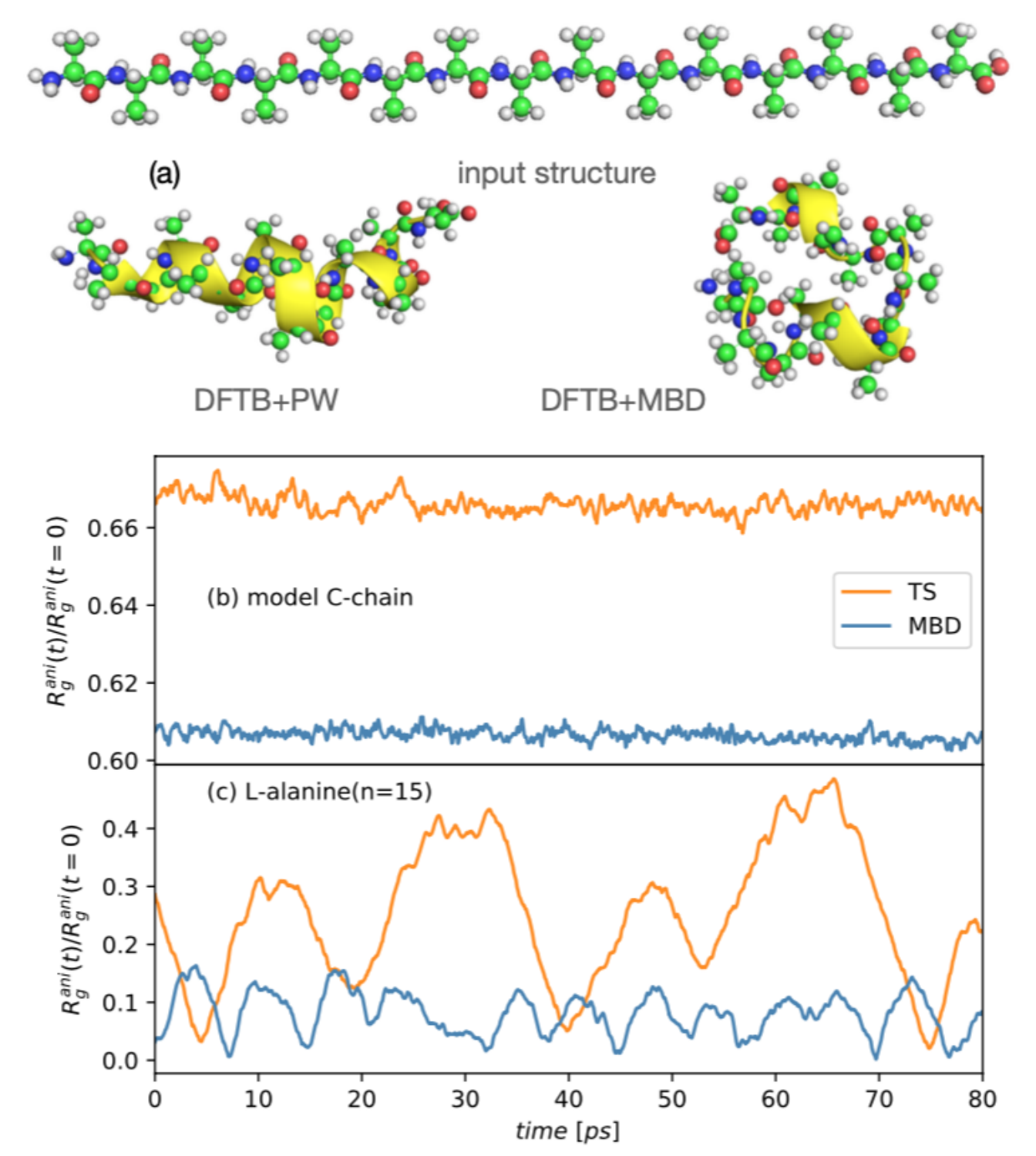}
	\caption{(Color Online) Top: Structure formation for poly-alanine (n=15).
	Time dependence of the anisotropy of the gyration ratio normalized to the initial geometries for a model Carbon chain (b) and for poly-alanine (c).
	We note that the absence of angular contributions in the covalent bond of the model implies a stronger stability to thermal fluctuations.
	On the contrary, the poly-alanine systems are simulated whithin a full treatment of the electronic structure, which therefore strongly compete with vdW interactions creating the oscillations displayed in panel (c).
	}
	\label{Fig4}
\end{figure}

As further validation, we present the results of \SI{300}{ps} simulations at \SI{300}{\kelvin} of a linear poly-alanine chain with 15 residues in gas phase.
At each time-step the electronic structure is treated within the DFTB+vdW method, which includes covalent bonding, electrostatics/polarization, Pauli repulsion and vdW interactions \cite{Hourahine2020}.
Despite the different simulation conditions, we find that the dynamical structure formation for each vdW approximation displays analogous features across the two systems.
For the model polymer, we find that the conformations corresponding to the lowest steady-state energies for PW has a regular helical shape.
However, the latter are asymmetric with respect to all three cartesian coordinates, which is not consistent with the cylindrical symmetry imposed by the constraint on the positions of the edge atoms.
On the contrary, MBD generally yields knot-like structures where the local regularity is sacrificed in favour of a cylindrical-like symmetry.
This originates in the versatility of MBD in addressing the anisotropy of the polarization response, which guides the conformational search towards geometries that are consistent with the geometrical constraints imposed by the environment.
Similar features are encountered in the case of the folding of the poly-alanine chain: Figure 4 (a) displays the conformation with the lowest steady-state energy (out of 5 runs per model) for PW and MBD.
Here no constraint on the chain edges was applied, and in both cases a local helical structure is obtained.
Nevertheless, in the PW case the chain maintains a globally linear structure while the MBD one folds on itself, optimizing the geometrical symmetry and resulting in a total energy gain of ~\SI{0.285}{\electronvolt}.
We remark that more compact, rather than seemingly linear, structures are consistent with previous studies on conformational searches lead by global non-specific interactions \cite{Osmanovic2016, Chakrabartty1995, Huyghues1995, Sung2017}. 
The obtained poly-alanine structures in the gas phase cannot be compared to protein structures formed under physiological conditions, which would require more sophisticated simulations in presence of solvent and ions.
Nevertheless, a force field trained on DFTB+MBD trajectories was recently applied to poly-alanine and crambin in solution with results consistent with experimental data and showing significant discrepancies with classical force fields \cite{Unke2022}.

We conclude by comparing the conformational stabilities at the steady-state for the lowest steady-state energy trajectories for each model and for the two systems.
Panels (b) and (c) of Fig.~\ref{Fig4} show the time-dependence of the anisotropy of the gyration ratio, normalized at the value for the initial structure, which measures the time evolution of the structural compactness.
On the one hand, the two vdW approximations yield distinctively different conformational anisotropy for the model polymer.
On the other hand, the conformational anisotropy of the poly-alanine has similar absolute minima, however the PW model yields oscillations with larger amplitude and lower frequency.
This indicates that the MBD conformations here observed are systematically more stable than the PW ones, in consistency with the higher entropic stability of MBD-optimized structures already demonstrated in the context of molecular crystals \cite{Reilly2014}.
The more pronounced instability of the poly-alanine systems can attributed to the competition between dispersion interactions and the angular component of the covalent bonding, that is not included in the model polymer.
Nevertheless, we note that the steady-state time-average of the anisotropy of the polarizability tensor in poly-alanine was calculated to be $\approx1.5$ times higher for PW rather than MBD, in full consistency with the discussion around Fig.~\ref{Fig2}.
In conclusion, we have shown that the inclusion of many-body effects in van der Waals dispersion interactions allows to account for anisotropic polarization responses, which depend on the global conformation.
Such effect yields an increased accuracy with respect to the PW approximation, however our results do not invalidate the numerous findings obtained neglecting many-body effects.
Nevertheless, our analysis shows that anisotropic dispersion interactions can be key to achieve conformations consistent with the geometrical constraints imposed by the environment, and future efforts will be dedicated to solvated proteins in connection to physiological conditions.


\begin{acknowledgments}
The authors acknowledge support from the ERC Consolidator Grant ``BeStMo".
MG is grateful to Marco Pezzutto and Matteo Gori for many fruitful discussions.
The calculations on the poly-alanine systems were carried out with the HPC facilities at the University of Luxembourg (see hpc.uni.lu).
\end{acknowledgments}

\bibliography{structure_formation}

\end{document}